# A Multi-Level Lorentzian Analysis of the Basic Structures of the Daily DJIA

**Frank W. K. Firk**

Professor Emeritus of Physics,

The Henry Koerner Center for Emeritus Faculty,

Yale University, New Haven , CT 06520

**Abstract**: In a recent communication (Firk, 2012), the use of a modified form of the Thomas theory of nuclear reactions as a model for the structures observed in the day trading of the DJIA, was introduced. Here, an analysis of the fine structure associated with High-Frequency Trading (HFT) and the intermediate structure (considered in two distinct components with differing adjacent spacing of the states) is carried out using time-dependent Lorentzian functions to model the fluctuations in the economic index. The parameters (the maximum value, the width (lifetime), and the time of each state) are obtained by fitting the data; they have statistical distributions that are closely related to those that are of fundamental importance in nuclear reaction theory. The key ratios $<\Delta\tau>/<D>$, where $<\Delta\tau>$ is the average lifetime, and $<D>$ is the average spacing between adjacent states (fluctuations), are determined for each structural form. In the case of the fine structure, the ratio $<\Delta\tau>/<D>$ is found to be remarkably constant throughout the trading day; this is a characteristic feature of the statistical nature of HFT. The value of this ratio in the first few hours of trading provides important information on the likely performance of the HFT component of the index for the remainder of the day.

1. **Introduction**

The striking similarity in the functional forms of the photoneutron cross sections observed in medium-mass nuclei (from magnesium to calcium), and the Dow Jones Industrial Average on a typical trading day, has been discussed in a previous paper (Firk, 2012). In particular, the breakdown of both the photonuclear cross section and the DJIA into basic fine, intermediate and gross structures was considered. In this paper, the fine and intermediate structures observed in the DJIA, on a particular day, are analyzed using time-dependent Lorentzian forms. (The Thomas approximation (Thomas, 1955) to the general R-matrix theory of nuclear reactions, discussed previously, contains many features that are more detailed than are required in the present analysis).

2. **A Time-Dependent Lorentzian Function**

The time-dependent Lorentzian function lzn(t) used here is conveniently written in terms of the variable x:

$$lzn(x) = M/(1 + x^2)$$

where

$$x = 2(t - t_0)/\Delta\tau,$$

$t_0$ is the center time of the symmetric function,

M is the maximum value of the function, and

$\Delta\tau$ is the full width at half maximum of the function.

The area, A, under the function, is obtained by integration; it is

$$A = \pi(M\Delta\tau)/2.$$

If the average adjacent spacing between states over an interval of time is <D> then the average value < lnz(t)> of the function lnz(t) is directly related to the ratio <Δτ>/<D>.

In the theory of nuclear reactions, a Lorentzian form is known as a Breit-Wigner form in which the variable is energy (Breit and Wigner, 1936). The width of a state is inversely related to the lifetime of the state. The exact theories of nuclear reactions (Kapur and Peierls, 1938, Wigner and Eisenbud, 1947 and Humblet and Rosenfeld, 1961) are developments of the Breit-Wigner theory.

**3. An Analysis of the DJIA data on April 3, 2012**

3.1. *Preparing the data*

We are concerned with a quantitative analysis of the highly structured form of the DJIA reported on April 3, 2012; the data are shown in 1-minute intervals.

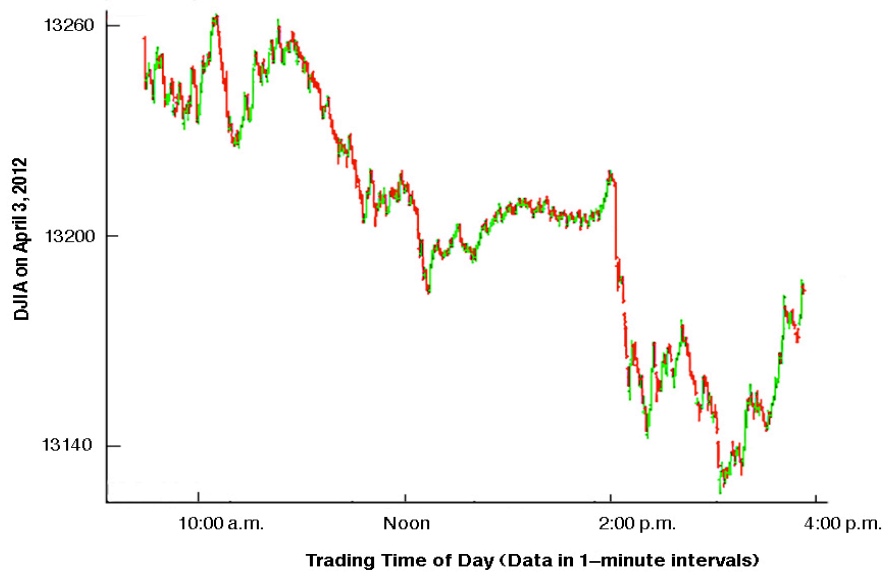

Figure 1. *The 1-minute interval data of the DJIA on April 3, 2012.*

In order to analyze the data using Lorentzians, it is necessary to understand the basic components of the economic index. These components form the backgrounds upon which the Lorentzian (resonant-like) states are based. The following forms of the underlying components are required; they are

1. Gross structure with a characteristic time-scale of many hours,

2. Intermediate structure that appears in two distinct patterns:

    I – with an average adjacent spacing of about one hour,

    II – with an average adjacent spacing of about 10 minutes,

3. Fine structure with a characteristic time-scale of 1- to 2-minutes. It is this component that is associated with High-Frequency Trading.

The gross and intermediate I components are illustrated:

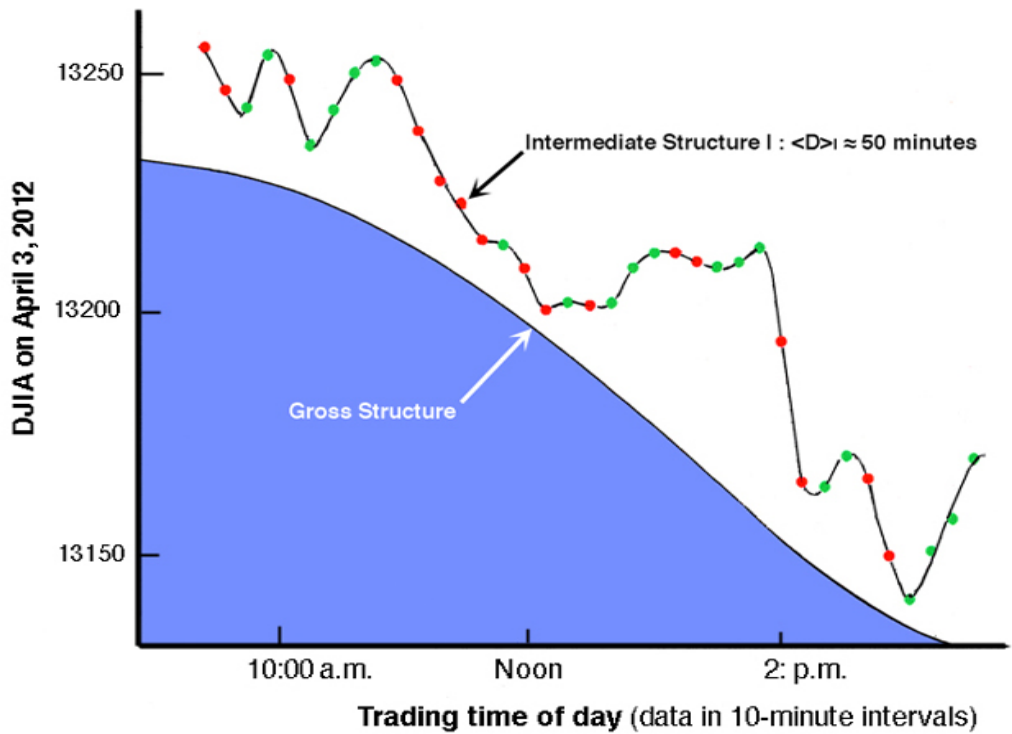

Figure 2. *The gross- and intermediate structure I obtained from the data in 10-minute intervals.*

The intermediate structure I, after subtraction of the gross structure, is shown in Figure 3.

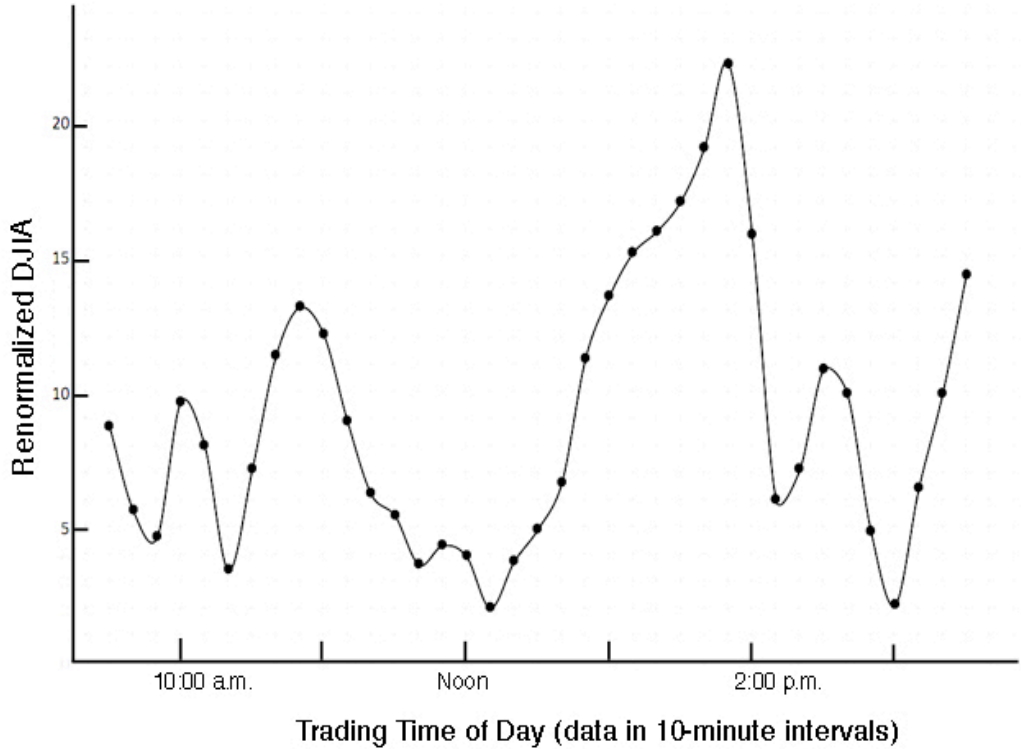

Figure 3. *The intermediate structure I after renormalization by the gross structure. The states have overlapping Lorentzian forms. The average spacing is $<D>_I \approx 50$ minutes.*

The process of renormalization is continued in order to obtain the Lorentzian forms of the intermediate structure II, and the fine structure. To achieve adequate time-resolution, the process requires data in 10-second-intervals; the individual points then can be averaged to give suitable statistical accuracy. In the present example, this is achieved by averaging in 30-second intervals. To illustrate the details of the analysis, the discussion is limited to a region of time of one hour.

The basic components – intermediate I, intermediate II, and fine structure – are summed to give the observed values of the DJIA in the time period 2:45 p.m. to 3:45 p.m.; they are shown in Figure 4.

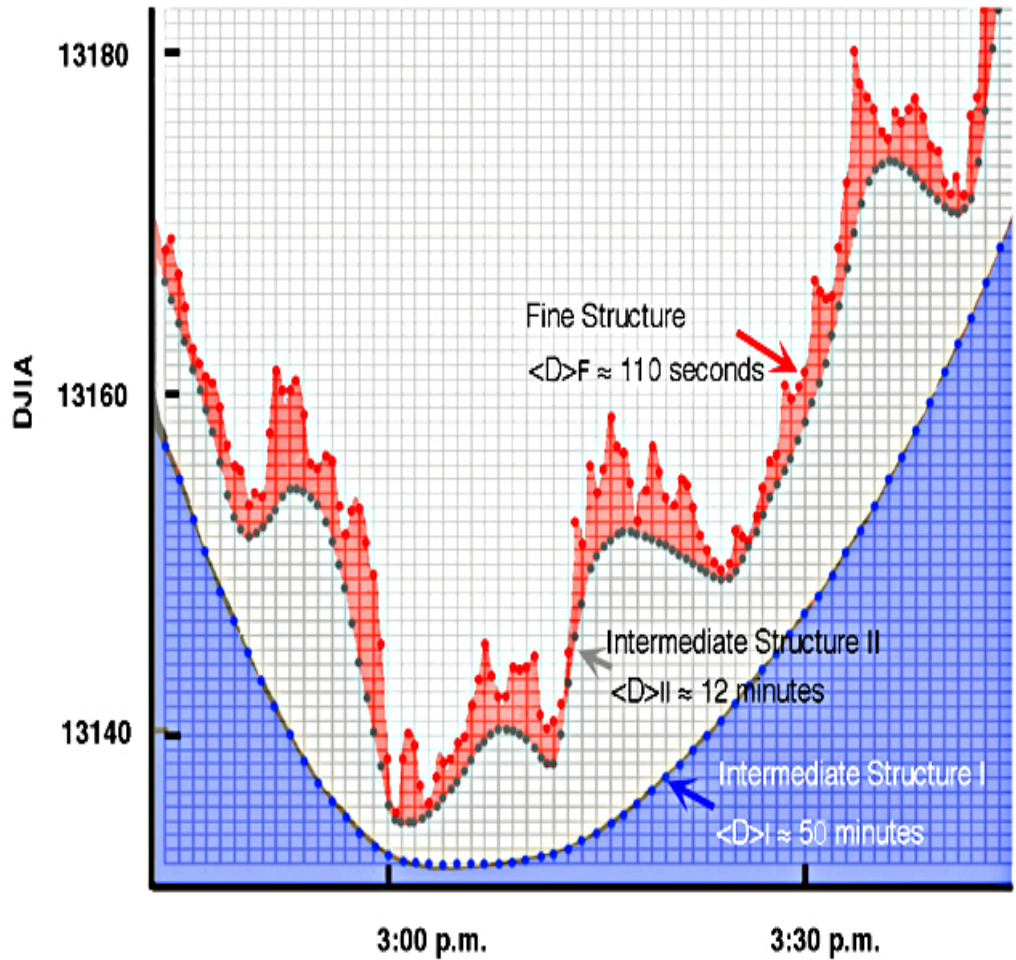

Figure 4. *Three components required in the analysis of the DJIA. The data are shown in 30-second intervals. The average spacing of adjacent states is $<D>_I \approx$ 50 minutes, $<D>_{II} \approx$ 12 minutes, and $<D>_F \approx$ 110 seconds.*

The renormalized fine structure is shown in Figure 5.

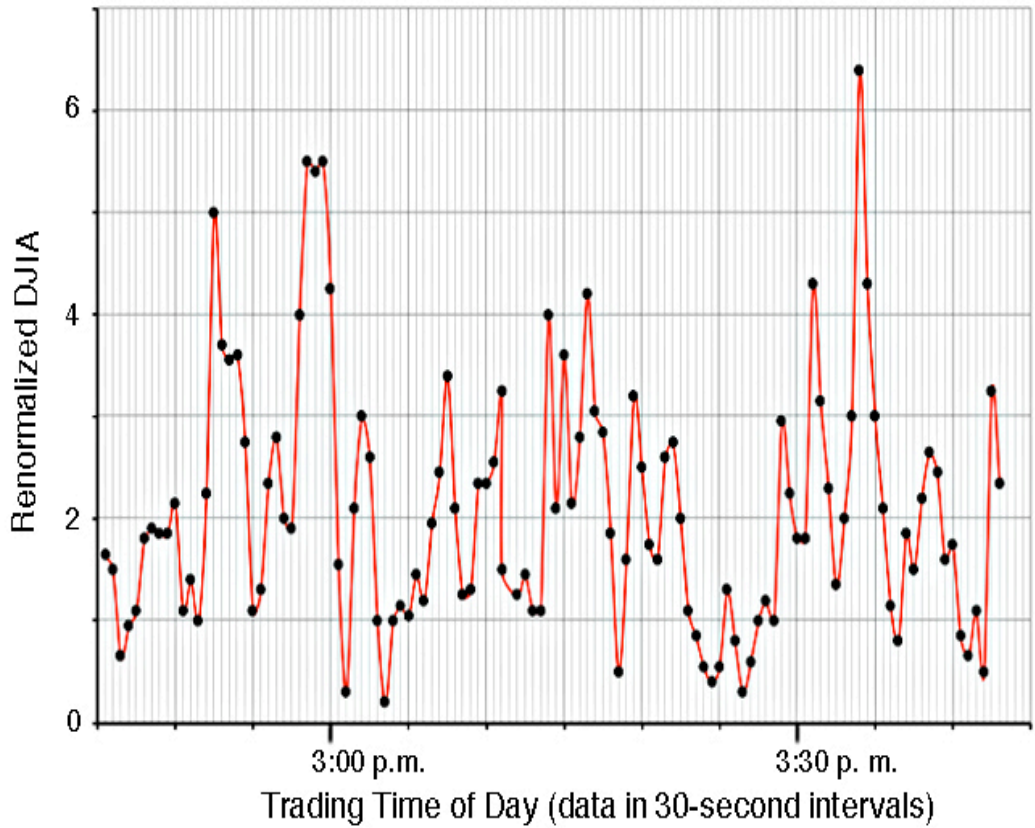

Figure 5. *The renormalized fine structure of the DJIA is shown in 30-second intervals. The average spacing is <D>F ≈ 110 seconds.*

The renormalized curves of the DJIA are now in forms that can be analyzed using a Lorentzian model.

3.1. *Intermediate Structure I*

An analytical fit to the 10-minute data is shown in Figure 6. An iterative method is used to analyze the overlapping states. The procedure is continued until a satisfactory fit is obtained; the small discrepancies that remain have a negligible effect on the value of the ratio <Δτ>/<D>.

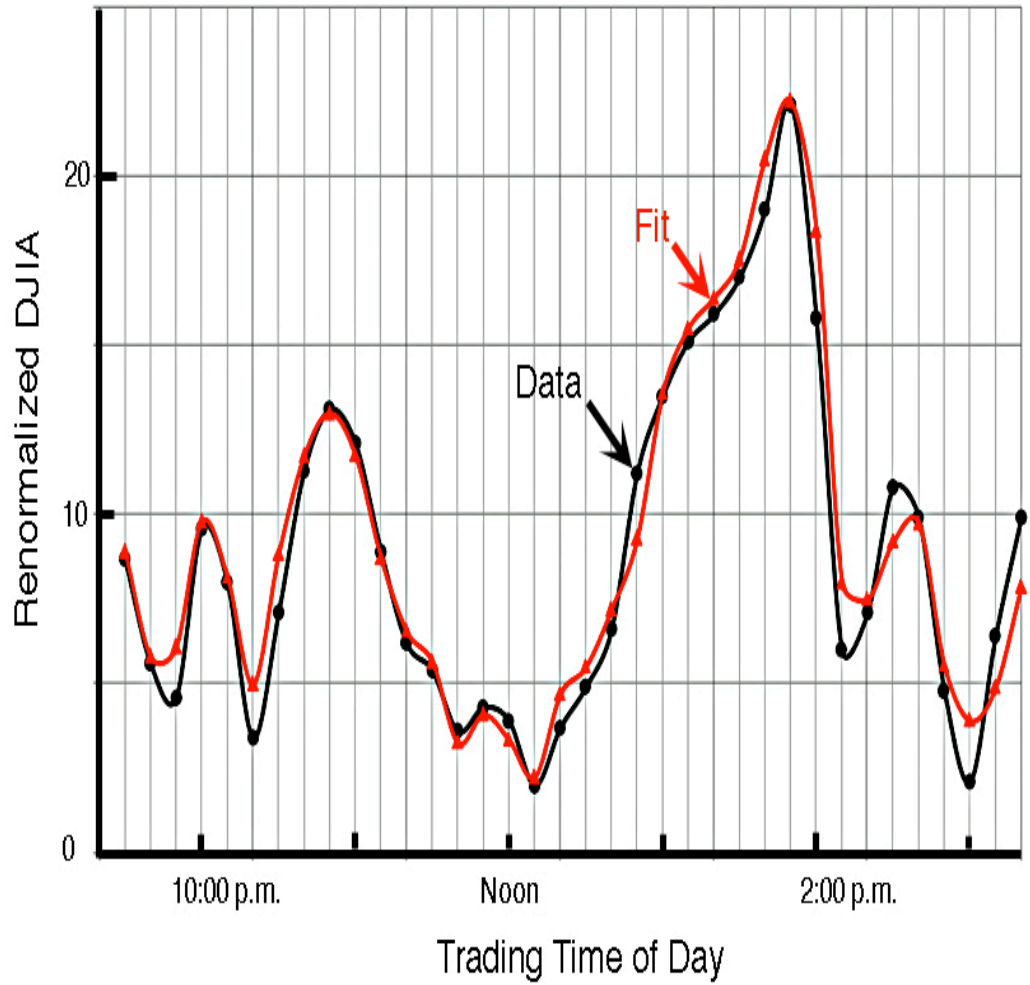

Figure 6. *Intermediate structure I; a multi-level Lorentzian analysis of the 10-minute data throughout the trading day.*

The value of $\langle\Delta\tau\rangle/\langle D\rangle \approx 0.6$; the small number of intermediate I states during the trading day limits the accuracy of the value.

3.2. *Intermediate Structure II*

An analytical fit to the 1-minute data in the time interval 2:45 p.m. to 3:45 p.m. is shown in Figure 7.

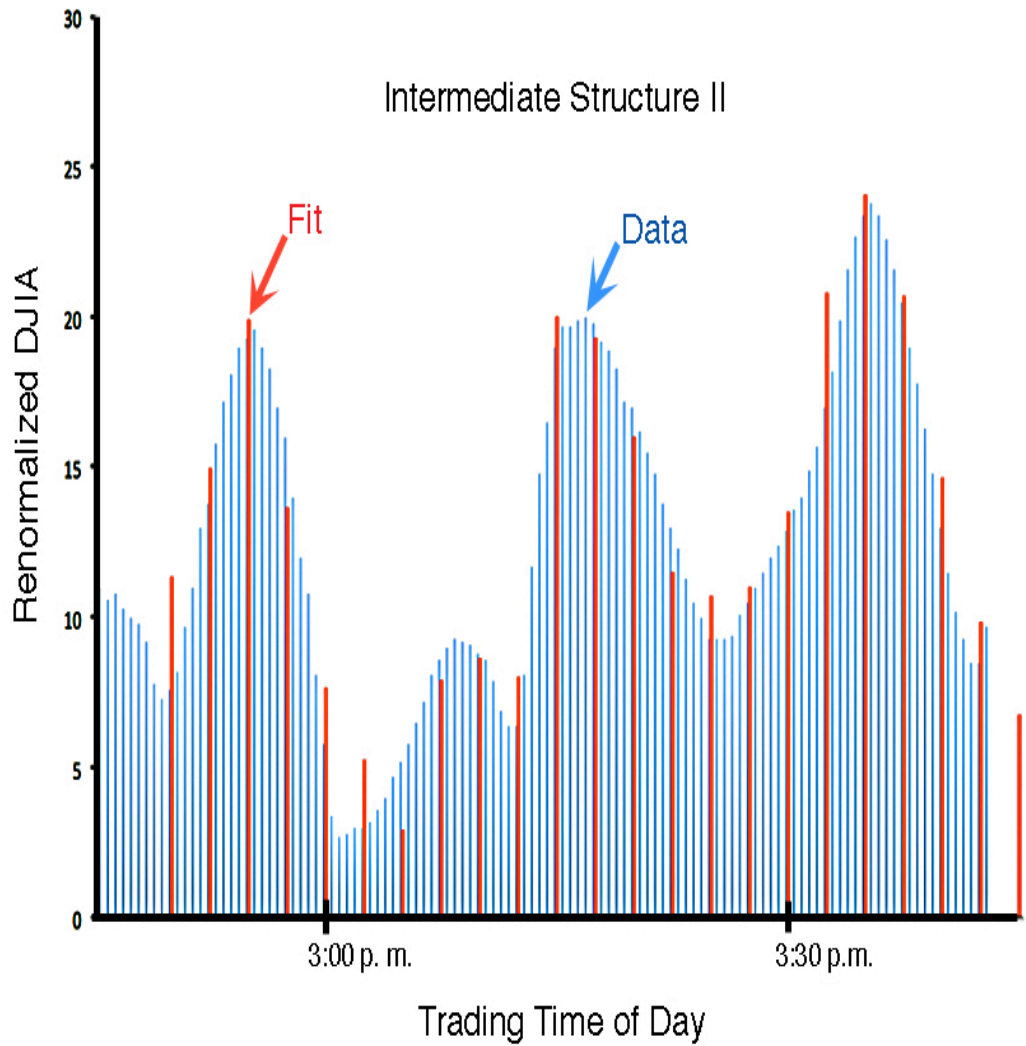

Figure 7. *A Lorentzian fit to the intermediate structure II during a 1-hour period. The data are shown in 30-second intervals, and the fit is shown in 150-second intervals.*

The value of $<D>_{II}$ for the intermediate II structure is obtained by averaging over the entire period of the trading day; it is $<D>_{II}$ = 12 minutes. The value of $<\Delta\tau>/<D>_{II} \approx 0.6$.

3.3 *The Fine Structure*

A multi-level fit to the fine structure data in the time interval 2:45 p.m. to 3:45 p.m. is shown in Figure 8.

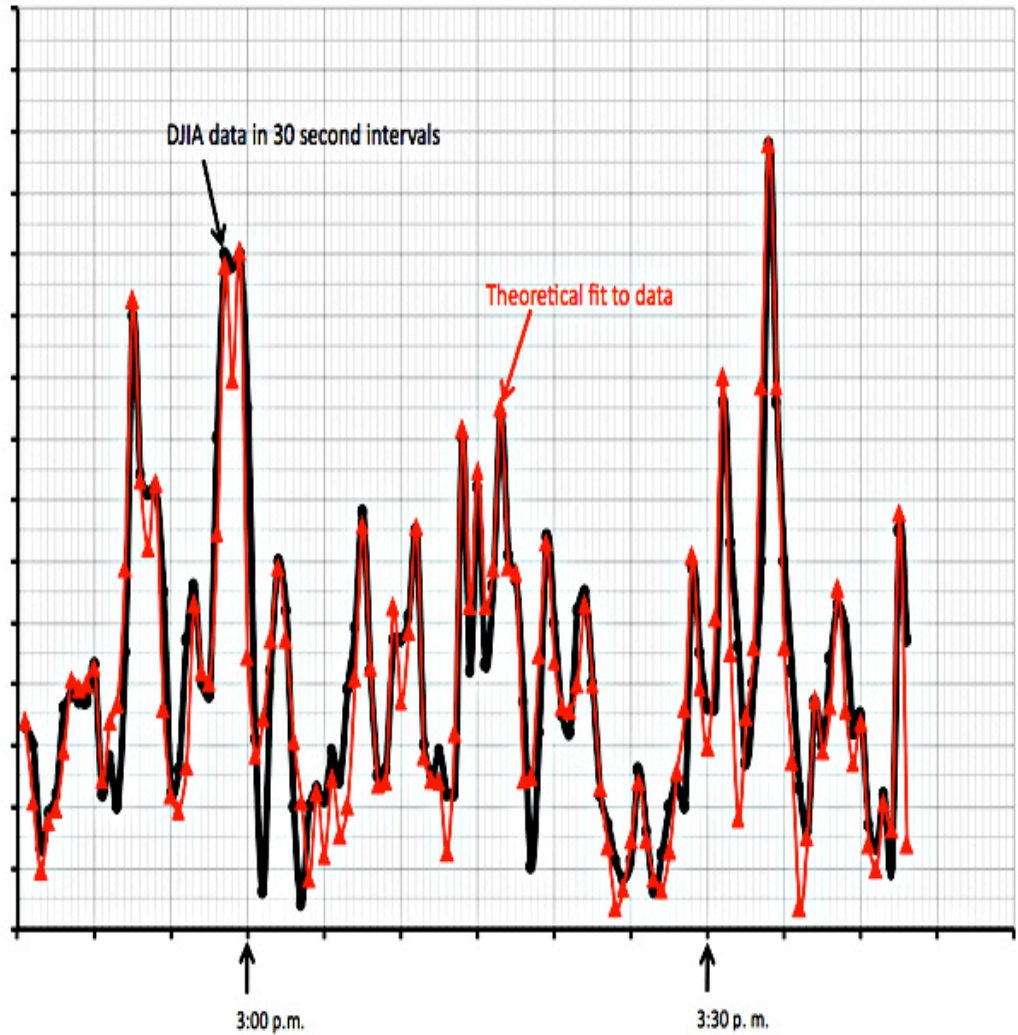

Figure 8. *A multi-level fit to the fine structure data from 2:45 p.m. to 3:45 p.m. is shown in red. The data are given in 30-second intervals.*

The parameters are listed in Table 3.

## Table 3

| Time of State Channel Number and time $t_0$, (seconds after 2:45 p. m.) | Width Number of Channels and $\Delta\tau$ (seconds) | Peak Height Number of Channels M, (renormalized DJIA) |
|---|---|---|
| 1  30 | 2.5  75 | 3 |
| 7  210 | 3  90 | 3.2 |
| 10  300 | 1  30 | 3.2 |
| 11  330 | 1  30 | 2.5 |
| 15  450 | 2.5  75 | 9 |
| 18  540 | 2  60 | 6.3 |
| 23  690 | 2  60 | 4.6 |
| 27  810 | 2.5  75 | 9 |
| 29  870 | 2  60 | 9 |
| 34  1020 | 2  60 | 5 |
| 39  1170 | 1.5  45 | 1.8 |
| 41  1230 | 1  30 | 2.2 |
| 45  1350 | 2.5  75 | 5.8 |
| 49  1470 | 1.5  45 | 3.8 |
| 51  1530 | 2  60 | 5.7 |
| 55  1650 | 0.7  21 | 1.5 |
| 58  1740 | 2  60 | 7 |
| 60  1800 | 1.5  45 | 6 |
| 63  1890 | 2.5  75 | 7.5 |
| 65  1950 | 1.5  45 | 3.7 |
| 69  2070 | 2  60 | 5.5 |
| 74  2220 | 2.5  75 | 4.5 |
| 81  2430 | 2  60 | 1.7 |
| 86  2580 | 2  60 | 1.4 |
| 88  2640 | 2  60 | 4.9 |
| 92  2760 | 2  60 | 9.8 |
| 98  2940 | 3  90 | 12 |
| 104  3120 | 1  30 | 1.7 |
| 107  3210 | 1.5  45 | 4.2 |
| 110  3300 | 1  30 | 2.5 |
| 113  3390 | 1  30 | 1.5 |
| 115  3450 | 2  60 | 5.8 |

The calculations are made using channel number units.

The average value of the adjacent spacing is $<D> = 110$ seconds, and the average value of the width is $<\Delta\tau> = 57$ seconds. The key ratio (the "strength function") is therefore

$$<\Delta\tau>/<D> = 0.52.$$

The individual states are in the resolved region and not in the region of Ericson fluctuations (Ericson, 1960), characterized by a value $<\Delta\tau>/<D> \gg 1$.

Within the limited statistical accuracy associated with the 32 states in the fine structure analysis, the width and spacing distributions are:

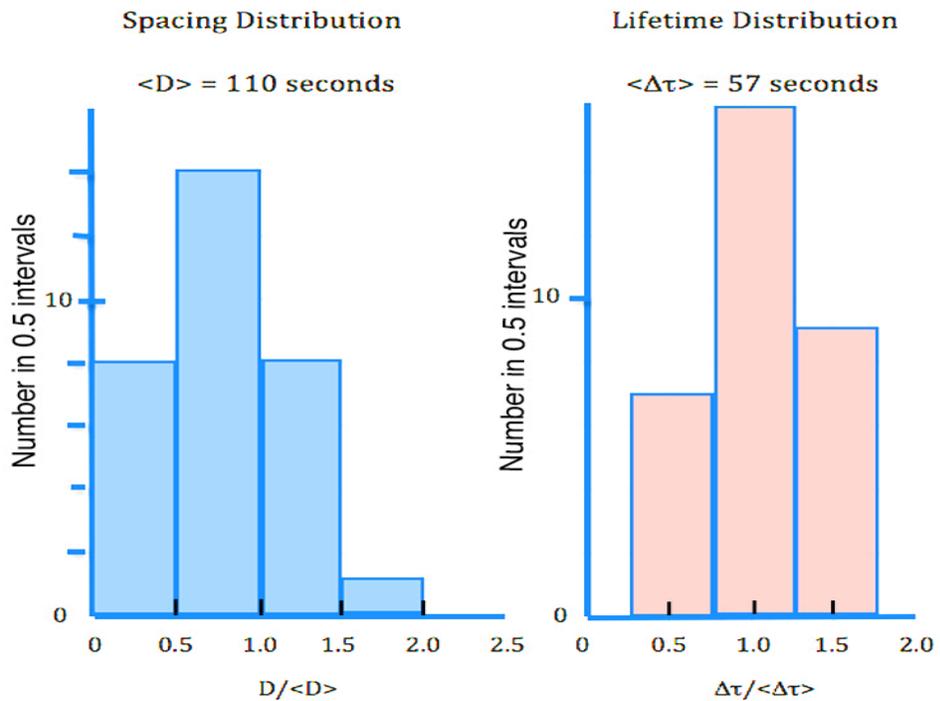

Figure 9. *The spacing and lifetime distributions associated with 32 fine structure states.*

These distributions have forms that are consistent with the underlying statistical nature of HFT; they are analogous to the statistical distributions of resonances observed in nuclear reactions in heavy nuclei. (See the discussion in Chapter 4).

The basic Lorentzian fine structure states, and their multi-level sum, are shown in Figure 10. The time interval covered is limited to about 15 minutes to illustrate the microscopic detail required in the analysis.

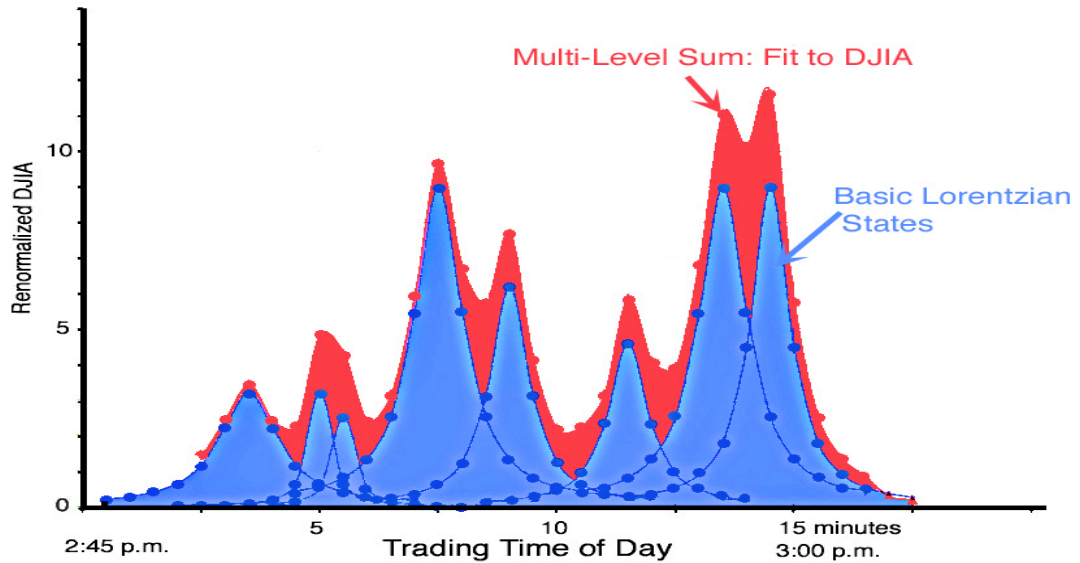

Figure 10. *The individual Lorentzian states that are summed to give the observed value of the DJIA in the time interval 2:45 p.m. 3:00 p.m.*

## 4. Statistical Basis of the Analysis

4.1. *Statistical Distributions in Nuclear Reactions*

The similarity in the form of the data of the DJIA shown in Figure 5, and the form of the total neutron cross section for the reaction neutron + 238-Uranium (Firk, Lynn and Moxon, 1963), shown in Figure 11, is noteworthy. The nuclear reaction is described in terms of Bohr's compound nucleus model (Bohr, 1936) that is statistical in its origin. The energy of an incoming nuclear particle is shared by all the interacting nucleons in the target nucleus, and a particle is emitted only after sufficient energy resides in one of the nucleons for it to escape from the nucleus. This is a statistical process, and takes a long time (on a nuclear time scale). The long lifetime is inversely related to the narrowness (on an energy scale) of the observed states by the Uncertainty Principle.

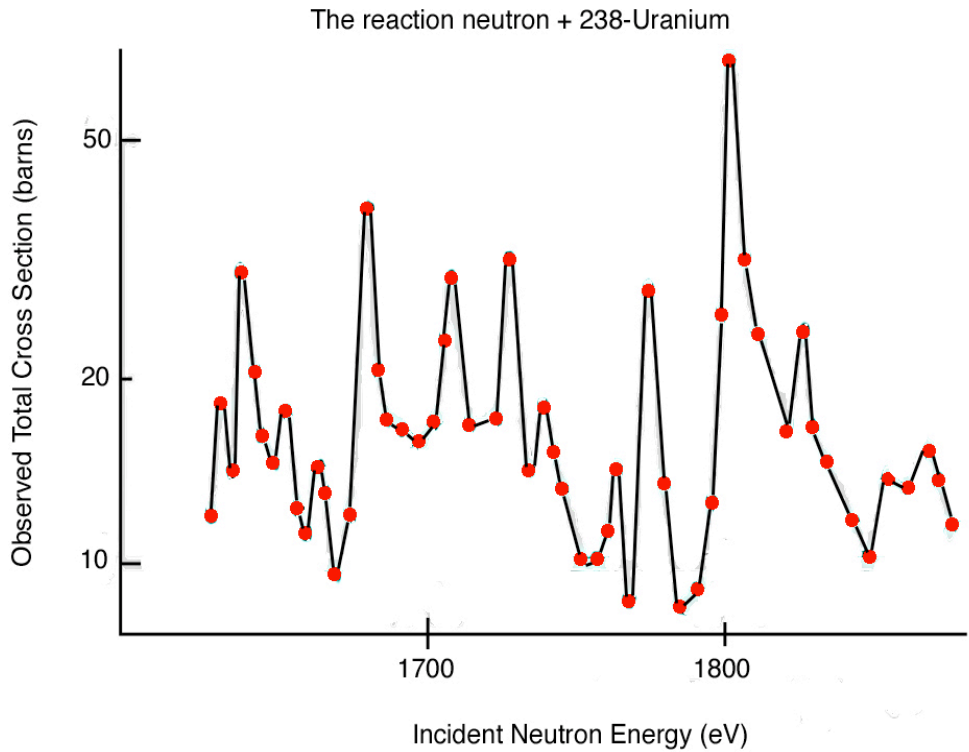

Figure 11. *The observed total neutron cross section for the interaction n + 238-Uranium (Firk, Lynn, and Moxon 1963).*

The spacing of adjacent resonances in nuclei with mass numbers A > 20 is found to be of a Wigner form (a Gaussian orthogonal ensemble (Wigner, 1956, Dyson, 1963)) the probability distribution function is

$$p(x) = (\pi/2) \cdot x \cdot \exp\{-\pi x^2/4\}$$

where x = spacing /average spacing.

This is a special case of the more general distribution, due to Weibull (1939); in its 2-parameter form, it is

$$\text{Wei}(x; b, c) = (c/b^c) \cdot x^{(c-1)} \cdot \exp\{-(x/b)^c\}.$$

The Wigner distribution is therefore

$$p(x) = \text{Wei}(x, 2/\sqrt{\pi}, 2).$$

The Wigner distribution has its origin in dealing with the complexity of the nuclear Hamiltonian that is generally infinitely dimensional, and has matrix elements that are unknown, except in a few cases associated with elastic scattering, and selected inelastic scattering. The matrix elements then are chosen randomly from some probability distribution function, usually a Gaussian. If time-reversal invariance is invoked, the theory leads to the important Gaussian Orthogonal Ensemble (Mehta, 2004).

The widths of the states belong to chi-squared distributions (Porter and Thomas, 1956).

$$\text{chi}(x; \nu) = \Gamma^{-1}(\nu/2)(\nu/2\langle x \rangle)^{\nu/2} \cdot x^{(\nu-2)/2} \cdot \exp\{-\nu x/2\langle x \rangle\}$$

where $\Gamma$ is the incomplete gamma function.

If $\nu = 2$, the Porter-Thomas distribution is obtained (for a single channel processes); it is an exponential. (The most probable width for elastic scattering is zero!).

For larger values of ν, the distribution goes to zero as x goes to zero: it is a more narrowly defined function, typical of a many-channel decay processes.

4.2. *Statistical Distributions of HFT*

The arguments used to justify the Wigner spacing distribution of adjacent states in compound nuclear reactions can, with suitably modified terminology, be used to justify the Wigner distribution of the fine structure states associated with HFT. The limited data set considered in section 3.3 is consistent with such a distribution. Previous work (Plerou et al., 2000) that covered an economic index over a period of years, found good agreement with a Wigner form for the longer-term spacing distribution. Effects of long-term correlations on the form of the distribution were also found.

The highly statistical nature of HFT is a direct consequence of the natural, well-defined response time of each trader, and the ultra high-speed flow of information now available. The narrow range of lifetimes of the HFT states is akin to the observed narrow range of lifetimes of nuclear resonances that decay by emitting radiation (gamma-rays); the total radiation width of a state is the sum of numerous partial radiation widths. The partial widths are associated with the many accessible states that are populated as the system gives up its total excitation energy in stages. For example, the average value of the *total* radiation widths of several hundred resonances in the reaction neutron + 238-Uranium is found to be $25 \pm 2$ meV. The numerous buying and selling options available to individual traders constitute a true many-channel, many-body system.

## 5. Conclusions.

An analysis of the DJIA on April 3, 2012 has been carried out using multi-level Lorentzian functions to model the index. In order to analyze the data it is necessary to understand, and to quantify, the basic structural components of the index. The structures that are necessary to represent the total index are 1) gross structures with time scales of many hours; 2) intermediate structures that appear in two distinct forms: i) structures with time scales of about one hour, and ii) structures with time scales of about 10 minutes; 3) fine structures with time scales of 1 to 2 minutes, associated with HFT.

The HFT states are excellent examples of discrete states with spacing and lifetime distributions that are well known, and well understood in the statistical models of low energy neutron resonance reactions (Porter, 1965). The parameters of the defining Lorentzians are used to obtain the fundamental ratio $<\Delta\tau>/<D>$. In the case of the HFT states studied, this ratio is about 0.5; it is found to be remarkably constant throughout the given trading day. This constancy is evidence for the purely statistical nature of HFT – it is unaffected by the uncontrollable, time-dependent market forces that can have such dramatic effects on the value of an economic index when studied over periods of many hours or days.

Within the statistics imposed by the limited number of states in the present study, the intermediate I and II structures have values $<\Delta\tau>/<D> \approx 0.6$. This indicates that the underlying states associated with these two structures are not Ericson fluctuations; they are states that also can be modeled using well-defined Lorentzians. More extensive analyses are required, however, to study in detail their statistical properties in order to uncover the possible effects of local correlations, and external market forces, on their structural forms.

**References**

Bohr N 1936 *Nature*, **137** 344

Dyson F J 1963 *J. Math. Phys.* **3** 140-156, 157-165, 166-175

Ericson T E O 1960 *Phys. Rev. Lett.* **5** 430

Firk F W K 2012 *arXiv.org:* **1203.6021v1**

Firk F W K, Lynn J E and Moxon M C 1963 *Nucl. Phys.* **41** 614

Humblet J and Rosenfeld L (1961) *Nucl. Phys.* **26** 529

Kapur P L and Peierls R (1938) *Proc. Royal Soc.* (London) **A 166** 277

Mehta M L 2004 *Random Matrices 3rd Edition* (San Diego CA: Elsevier Inc)

Plerou V, Gopikrishnan P, Rosenow B, Amaral L A N, and Stanley H E 2000 *Physica* **A 287** 374

Porter C E 1965 *Statistical Theories of Spectra: Fluctuations* (New York: Academic Press)

Porter C E and Thomas R G 1956 *Phys. Rev.* **104** 483

Thomas R G 1955 *Phys. Rev.* **97** 224

Wigner E P and Eisenbud L (1947) *Phys. Rev.* **72** 29

Wigner E P 1957 *Proc. Conf. on Neutron Physics by Time-of-Flight Methods* (Gatlinburg TN, Nov 1956) Oak Ridge National Lab. Report ORNL-2309